\documentclass[12pt]{iopart}

\usepackage{graphicx}

\begin{document}

\title[Wave propagation in stratified bi-isotropic media]{Invariant imbedding theory of wave propagation in arbitrarily inhomogeneous stratified bi-isotropic media}

\author{Seulong Kim and Kihong Kim}

\address{Department of Energy Systems
Research and Department of Physics, Ajou University, Suwon 16499, Korea}

\ead{khkim@ajou.ac.kr}

\begin{abstract}
Bi-isotropic media, which include isotropic chiral media
and Tellegen media as special cases, are the most general form of linear isotropic media where the electric displacement and the magnetic induction
are related to both the electric field and the magnetic intensity. In inhomogeneous bi-isotropic media, electromagnetic waves of two different polarizations are coupled to each other.
In this paper, we develop a generalized version of the invariant imbedding method for the study of wave propagation in arbitrarily-inhomogeneous stratified bi-isotropic media, which can be used to solve the coupled wave propagation problem accurately and efficiently. We verify the validity and usefulness of the method by applying it to several examples, including the wave propagation in a uniform chiral slab, the surface wave excitation in a bilayer system made of a layer of Tellegen medium and a metal layer, and the mode conversion of transverse electromagnetic waves into longitudinal plasma oscillations in inhomogeneous Tellegen media. In contrast to the case of ordinary isotropic media, we find that the surface wave excitation and the mode conversion occur for both $s$ and $p$ waves in bi-isotropic media.
\end{abstract}

\maketitle


\section{Introduction}

Bi-isotropic media, which include isotropic chiral media
and Tellegen media as special cases, are the most general form of linear isotropic media,
where the electric displacement $\bf D$ and the magnetic induction $\bf B$
are related to both the electric field $\bf E$ and the magnetic intensity $\bf H$ \cite{1}.
The unusual constitutive relations of these media are the consequences of the
magnetoelectric (ME) effects occurring in ME materials or ME metamaterials, which are due to
the coupling between electric polarization and magnetization \cite{2}.
In those materials, an electric field can induce magnetization and a magnetic field can induce electric polarization.
The study of ME materials has a long history going back to the works of
Curie, Debye, Landau and Lifshitz, Dzyaloshinskii, and Astrov \cite{2,3}.
The research of various materials showing strong ME effects including multiferroics is currently
a very active area of investigation in condensed matter physics \cite{3}.

The bi-isotropic medium is the name given to a magnetoelectric material, when one is especially interested
in the properties associated with electromagnetic wave propagation.
Two different kinds of bi-isotropic media, which have distinct ways of magnetoelectric coupling, have been studied in detail. The more extensively studied case is that of isotropic chiral media, in which the mirror symmetry is broken \cite{4}. The other kind is called Tellegen medium, in which the Lorentz reciprocity theorem is not obeyed \cite{1,5,6,7}.
Recently, a large amount of attention has been given to chiral metamaterials with a strong chirality \cite{8,9,10,11,12,13} and to topological insulators \cite{14,15,16,17,18}. In the presence of weak time-reversal-symmetry-breaking perturbations, topological insulators can be considered as a kind of Tellegen media.

In spatially uniform bi-isotropic media, right-circularly-polarized (RCP) and left-circularly-polarized (LCP) waves are the eigenmodes of propagation \cite{1}. In isotropic chiral media, the effective refractive indices associated with RCP and LCP waves are different, whereas the impedance is independent of the helicity. This gives rise to circular birefringence phenomena. In isotropic chiral media with a non-uniform impedance distribution, RCP and LCP waves are no longer eigenmodes and are coupled to each other. In contrast, in Tellegen media, the effective impedances associated with RCP and LCP waves are different, whereas the effective refractive index is the same for both waves. Again, these waves are coupled to each other in spatially non-uniform media.

In order to investigate the wave propagation characteristics in inhomogeneous bi-isotropic media, it is necessary to have a theoretical method to analyze the propagation of two coupled waves in such media. A generalized version of the invariant imbedding method \cite{19,20,21,22,23}, which can analyze the propagation of any number of coupled waves in arbitrarily inhomogeneous stratified media accurately and efficiently, has been developed in \cite{24} and applied to isotropic chiral media \cite{13,24}.
In this paper, we generalize this method further to more general situations and use it to study wave propagation in stratified bi-isotropic media.
We apply our method to several examples, including the surface wave excitation at the interface between a Tellegen medium and a metal and the mode conversion of transverse electromagnetic waves into longitudinal plasma oscillations occurring in inhomogeneous Tellegen media. The main aim of these examples is not to give an extensive analysis of these phenomena, but to verify the validity of our invariant imbedding method and to demonstrate its usefulness and efficiency.

In section \ref{sec2}, we derive the coupled wave equations in stratified bi-isotropic media. In section \ref{sec3} and appendix A, we describe the invariant imbedding method used in this paper and derive the invariant imbedding equations. In section \ref{sec4}, we elaborate on the expressions for the reflection and transmission coefficients and demonstrate that in lossless cases, the energy conservation law is properly obeyed. In section \ref{sec5}, we show the results of numerical calculations for a uniform chiral slab, the surface wave excitation at the interface between a Tellegen medium and a metal and the mode conversion occurring in inhomogeneous Tellegen media. Finally, in section \ref{sec6}, we give a summary of the paper.

\section{Coupled wave equations in stratified bi-isotropic media}
\label{sec2}

In this paper, we will use cgs Gaussian units.
In linear bi-isotropic media, the constitutive relations for harmonic waves are given by
\begin{eqnarray}
&&{\bf D}=\epsilon{\bf E}+a{\bf H},\nonumber\\
&&{\bf B}=\mu{\bf H}+a^*{\bf E},
\end{eqnarray}
where $\epsilon$ is the dielectric permittivity and $\mu$ is the magnetic permeability.
The complex-valued magnetoelectric parameter $a$ is expressed as
\begin{equation}
a=\chi+i\gamma,
\end{equation}
where $\chi$ is the non-reciprocity (or Tellegen) parameter and $\gamma$ is the chirality index.
In bi-isotropic media, $\epsilon$, $\mu$, $\chi$ and $\gamma$ are dimensionless scalar quantities.
When $\epsilon$ and $\mu$ are real, the bi-isotropic medium is lossless. In the case of reciprocal chiral media,
these constitutive relations have been argued to arise from the effects of weak spatial dispersion \cite{25,26}.

From Maxwell's curl equations and the constitutive relations, we derive the
wave equations satisfied by the electric and magnetic fields in inhomogeneous bi-isotropic media for harmonic waves:
\begin{eqnarray}
\mu\nabla\times\left(\frac{\nabla\times{\bf E}}{\mu}\right)&=&k_0^2\left(\epsilon\mu-\vert a\vert^2\right){\bf E}+ik_0\left(a^*-a\right)\nabla\times{\bf E}\nonumber\\&&+ik_0\mu\nabla\left(\frac{{a^*}}{\mu}\right)\times{\bf E},\nonumber\\
\epsilon\nabla\times\left(\frac{\nabla\times{\bf H}}{\epsilon}\right)&=&k_0^2\left(\epsilon\mu-\vert a\vert^2\right){\bf H}+ik_0\left(a^*-a\right)\nabla\times{\bf H}\nonumber\\&&-ik_0\epsilon\nabla\left(\frac{a}{\epsilon}\right)\times{\bf H},
\label{eq:weq}
\end{eqnarray}
where $k_0$ ($=\omega/c$) is the vacuum wave number. When $\chi$ is zero, the wave equation for ${\bf E}$ reduces to (35) of \cite{4}.
We note that there is a symmetry between these two equations under the transformation ${\bf E}\rightarrow{\bf H}$,
${\bf H}\rightarrow -{\bf E}$, $\epsilon\rightarrow \mu$, $\mu\rightarrow \epsilon$ and $a\rightarrow -a^*$.
We restrict ourselves to stratified media where the parameters $\epsilon$, $\mu$, $\chi$ and $\gamma$
depend only on one spatial coordinate, $z$.
For plane waves propagating in the $xz$ plane, the $x$ dependence
of all field components is contained in the factor $e^{iqx}$, where $q$ is the $x$ component of the wave vector.
Then we can eliminate $E_x$, $E_z$, $H_x$ and $H_z$
from (\ref{eq:weq}) and obtain two coupled wave equations
satisfied by $E_y=E_y(z)$ and $H_y=H_y(z)$:
\begin{eqnarray}
{{d^2 \psi}\over{dz^2}}-\frac{d\cal E}{dz}{\cal E}^{-1}(z)
\frac{d\psi}{dz}+\left[k_0^2{\cal E}(z){\cal M}(z)-q^2I\right]\psi=0,
\label{eq:wave}
\end{eqnarray}
where $I$ is a $2\times 2$ unit matrix and
\begin{eqnarray}
\psi=\pmatrix{ E_y \cr H_y \cr},~~{\cal E}=\pmatrix{\mu&-a^*\cr -a&\epsilon\cr},~~
{\cal M}=\pmatrix{\epsilon&a\cr a^*&\mu\cr}.
\label{eq:matrix}
\end{eqnarray}
After obtaining $E_y$ and $H_y$ by solving (\ref{eq:wave}), we can calculate the
field components $E_x$, $H_x$, $E_z$ and $H_z$ using the relationships
\begin{eqnarray}
&&E_x=-\frac{i}{k_0}\frac{1}{\epsilon\mu-\vert a\vert^2}\left(a\frac{dE_y}{dz}+\mu\frac{dH_y}{dz}\right),\nonumber\\
&&H_x=\frac{i}{k_0}\frac{1}{\epsilon\mu-\vert a\vert^2}\left(\epsilon\frac{dE_y}{dz}+a^*\frac{dH_y}{dz}\right),\nonumber\\
&&E_z=-\frac{q}{k_0}\frac{1}{\epsilon\mu-\vert a\vert^2}\left(a E_y+\mu H_y\right),\nonumber\\
&&H_z=\frac{q}{k_0}\frac{1}{\epsilon\mu-\vert a\vert^2}\left(\epsilon E_y+a^* H_y\right).
\label{eq:fc}
\end{eqnarray}

\section{Invariant imbedding equations}
\label{sec3}

We assume that the waves are incident from a uniform dielectric region ($z>L$)
where $\epsilon=\epsilon_1$, $\mu=\mu_1$ and $a=0$ and
transmitted to another uniform dielectric region ($z<0$)
where $\epsilon=\epsilon_2$, $\mu=\mu_2$ and $a=0$. We emphasize that the incident and transmitted regions are filled with ordinary dielectric
media with zero magnetoelectric parameter.
The inhomogeneous bi-isotropic medium
of thickness $L$ lies in $0\le z\le L$.
In the special case where the waves are incident from a vacuum with $\epsilon_1=\mu_1=1$,
the invariant imbedding equations for the reflection coefficient, the transmission coefficient and the electromagnetic fields
corresponding to (\ref{eq:wave}) have been derived previously in \cite{24}.
There is, however, some ambiguity in generalizing that result to the case where $\epsilon_1$ and $\mu_1$ take arbitrary values
by using the method in \cite{24}.
In this paper, we develop a different method of deriving invariant imbedding equations, using which
we achieve the aforementioned generalization. This method also has an advantage in that it can be used to solve coupled wave equations that are much more general than (\ref{eq:wave}).

We generalize
(\ref{eq:wave}) by replacing the vector wave function
$\psi$ by a $2\times 2$ matrix wave function $\Psi$, the $j$-th
column vector $(\Psi_{1j},\Psi_{2j})^T$ of which represents
the wave function when the incident wave consists only of the $j$-th
wave ($j=1,2$). We note that the index $j=1$ ($j=2$) corresponds to the case where $s$ ($p$) waves are incident.
We are interested in calculating the $2\times 2$ reflection and
transmission coefficient matrices $r=r(L)$ and $t=t(L)$, which we consider as functions of $L$.
In our notation,
$r_{21}$ is the reflection coefficient
when the incident wave is $s$-polarized and the reflected wave is $p$-polarized.
Similarly, $r_{12}$ is the reflection coefficient
when the incident wave is $p$-polarized and the reflected wave is $s$-polarized.
Similar definitions are applied to the transmission coefficients.

It is convenient to rewrite the wave equation in the form
\begin{eqnarray}
\left({\cal E}^{-1}\Psi^\prime\right)^\prime+{\cal D}\Psi=0,~~~{\cal D}=k_0^2{\cal M}-q^2{\cal E}^{-1},
\label{eq:wave2}
\end{eqnarray}
where the prime denotes a differentiation with respect to $z$. We introduce $2\times 2$ matrix functions
\begin{eqnarray}
U_1(z;L)=\Psi,~~~U_2(z;L)={\cal E}^{-1}\Psi^\prime,
\end{eqnarray}
which we consider as functions of both $z$ and $L$. Then the wave equation is transformed to
\begin{eqnarray}
\pmatrix{U_1 \cr U_2}^\prime=A\pmatrix{U_1 \cr U_2},~~~A=\pmatrix{ O & \mathcal{E} \cr -\mathcal{D} & O},
\end{eqnarray}
where $A$ is a $4\times 4$ matrix and $O$ is a $2\times 2$ null matrix.

The wave functions in the incident and transmitted regions are expressed in
terms of $r$ and $t$:
\begin{eqnarray}
U_1(z;L)=\left\{ \begin{array}{ll} e^{ip(L-z)}I +e^{ip(z-L)}~r,
&~z>L\\
e^{-ip^\prime z}~t, &~z<0 \end{array} \right., \label{eq:psi}
\end{eqnarray}
where $p$ and $p^\prime$ are the {\it negative} $z$ components of the wave vector
in the incident and transmitted regions.
When $\epsilon_1\mu_1$ and $\epsilon_2\mu_2$ are positive real numbers,
$p$ and $p^\prime$ are obtained from the relationships
\begin{eqnarray}
p^2+q^2=k_0^2\epsilon_1\mu_1,~~~{p^{\prime}}^2  +q^2=k_0^2\epsilon_2\mu_2.
\end{eqnarray}
If $\theta$ is the incident angle, $p$, $q$ and $p^\prime$ are expressed as
\begin{eqnarray}
p&=&k_0\sqrt{\epsilon_1\mu_1}\cos\theta,~~~
q=k_0\sqrt{\epsilon_1\mu_1}\sin\theta,\nonumber\\
p^\prime&=&\left\{ \begin{array}{ll}
k_0\sqrt{\epsilon_2\mu_2-\epsilon_1\mu_1\sin^2\theta},  &  ~\epsilon_2\mu_2\ge\epsilon_1\mu_1\sin^2\theta \\
ik_0\sqrt{\epsilon_1\mu_1\sin^2\theta-\epsilon_2\mu_2},  &  ~\epsilon_2\mu_2<\epsilon_1\mu_1\sin^2\theta  \end{array} \right..
\label{eq:pqr}
\end{eqnarray}
At the boundaries of the inhomogeneous medium, we have
\begin{eqnarray}
&&U_1(0;L)=t,~~~U_1(L;L)=r+I,\nonumber\\
&&U_2(0;L)=-ip^\prime{\mathcal{E}}_2^{-1}t=-ip^\prime{\mathcal{E}}_2^{-1}U_1(0;L),\nonumber\\
&&U_2(L;L)=ip{\mathcal{E}}_1^{-1}\left(r-I\right)=ip{\mathcal{E}}_1^{-1}U_1(L;L)-2ip\mathcal{E}^{-1}_1,
\label{eq:bcz}
\end{eqnarray}
where ${\cal E}_1$ and ${\cal E}_2$ are the values of $\cal E$ in the incident and transmitted regions respectively.
From (\ref{eq:bcz}), we get
\begin{eqnarray}
g{\hat S}+h{\hat R}=v,
\end{eqnarray}
where
\begin{eqnarray}
&&{\hat S}=\pmatrix{U_1(0;L) \cr U_2(0;L)}, ~~~{\hat R}=\pmatrix{U_1(L;L) \cr U_2(L;L)},\nonumber\\
&&g=\pmatrix{ip^\prime{\mathcal{E}}_2^{-1} & I \cr O & O},~~~
h=\pmatrix{ O & O \cr ip{\mathcal{E}}_1^{-1} & -I},~~~
v=\pmatrix{ O \cr 2ip{\mathcal{E}}_1^{-1}}.
\end{eqnarray}
We define $4 \times 4$ matrices $\mathcal S$ and $\mathcal R$ by
\begin{eqnarray}
{\hat S}={\mathcal S}v=\pmatrix{S_{11} & S_{12} \cr S_{21} & S_{22}}v,~~~{\hat R}={\mathcal R}v=\pmatrix{R_{11} & R_{12} \cr R_{21} & R_{22}}v,
\end{eqnarray}
where $S_{ij}$ and $R_{ij}$ ($i,j=1,2$) are $2\times 2$ matrices.
The invariant imbedding equations satisfied by ${\mathcal R}$ and ${\mathcal S}$ have been derived in appendix A:
\begin{eqnarray}
&&\frac{d{\mathcal R}}{dl}=A(l){\mathcal R}(l)-{\mathcal R}(l)h A(l){\mathcal R}(l),\nonumber\\
&&\frac{d{\mathcal S}}{dl}=-{\mathcal S}(l)h A(l){\mathcal R}(l),
\label{eq:iea}
\end{eqnarray}
where $l$ is the thickness of the inhomogeneous layer in the $z$ direction \cite{27,28}.
The initial conditions for these matrices are given by
\begin{eqnarray}
{\mathcal R}(0)={\mathcal S}(0)=\left(g+h\right)^{-1}.
\label{eq:icaa}
\end{eqnarray}
From the definitions of $\hat R$, $\hat S$, $\mathcal R$ and $\mathcal S$, we obtain
\begin{eqnarray}
&&R_{12}=\frac{1}{2ip}\left(r+I\right)\mathcal{E}_1,~~
R_{22}=\frac{1}{2}\left(\mathcal{E}^{-1}_1 r\mathcal{E}_1-I\right),~~
S_{12}=\frac{1}{2ip}t\mathcal{E}_1.
\label{eq:ieb}
\end{eqnarray}

The invariant imbedding equations satisfied by $r$ and $t$ follow from (\ref{eq:iea}) and (\ref{eq:ieb})
and take the forms
\begin{eqnarray}
&&\frac{dr}{dl}=ip\left(r\mathcal{E}\mathcal{E}_1^{-1}+\mathcal{E}\mathcal{E}_1^{-1}r\right)
-\frac{ip}{2}\left(r+I\right)\left(\mathcal{E}\mathcal{E}^{-1}_1-\frac{1}{p^2}\mathcal{E}_1\mathcal{D}\right)\left(r+I\right),\nonumber\\
&&\frac{dt}{dl}=ipt\mathcal{E}\mathcal{E}_1^{-1}
-\frac{ip}{2}t\left(\mathcal{E}\mathcal{E}_1^{-1}-\frac{1}{p^2}\mathcal{E}_1\mathcal{D}\right)\left(r+I\right).
\end{eqnarray}
By substituting the expression for $\cal D$ into this equation and using the identity
$k_0^2\epsilon_1\mu_1=p^2+q^2$,
we finally obtain
\begin{eqnarray}
&&\frac{dr}{dl}=ip\left(r{\tilde\mathcal{E}}+{\tilde\mathcal{E}}r\right)
+\frac{ip}{2}\left(r+I\right)\left[{\tilde\mathcal{M}}-{\tilde\mathcal{E}}+\frac{q^2}{p^2}\left({\tilde{\mathcal M}}-{\tilde{\mathcal E}}^{-1}\right)\right]\left(r+I\right),\nonumber\\
&&\frac{dt}{dl}=ipt{\tilde\mathcal{E}}
+\frac{ip}{2}t\left[{\tilde\mathcal{M}}-{\tilde\mathcal{E}}+\frac{q^2}{p^2}\left({\tilde{\mathcal M}}-{\tilde{\mathcal E}}^{-1}\right)\right]\left(r+I\right),
\label{eq:iec}
\end{eqnarray}
where $\tilde\mathcal{E}$ and $\tilde\mathcal{M}$ are defined by
\begin{eqnarray}
&&{\tilde \mathcal E}\equiv{\mathcal E}{\mathcal E}_1^{-1}=\pmatrix{\frac{\mu}{\mu_1}&-\frac{a^*}{\epsilon_1}\cr -\frac{a}{\mu_1}&\frac{\epsilon}{\epsilon_1}\cr}=\pmatrix{\frac{\mu}{\mu_1}&\frac{-\chi+i\gamma}{\epsilon_1}\cr \frac{-\chi-i\gamma}{\mu_1}&\frac{\epsilon}{\epsilon_1}\cr},\nonumber\\
&&{\tilde \mathcal M}\equiv\frac{1}{\epsilon_1\mu_1}{\mathcal E}_1{\mathcal M}=\pmatrix{\frac{\epsilon}{\epsilon_1}&\frac{a}{\epsilon_1}\cr \frac{a^*}{\mu_1}&\frac{\mu}{\mu_1}\cr}=\pmatrix{\frac{\epsilon}{\epsilon_1}&\frac{\chi+i\gamma}{\epsilon_1}\cr \frac{\chi-i\gamma}{\mu_1}&\frac{\mu}{\mu_1}\cr}.
\label{eq:matrix2}
\end{eqnarray}
When $\epsilon_1=\mu_1=1$ and $a=i\gamma$, these equations have the precisely same form as those presented in \cite{24}.

The initial conditions for $r$ and $t$ are obtained from (\ref{eq:icaa}):
\begin{eqnarray}
&&r_{11}(0)=\frac{\mu_2p-\mu_1 p^\prime}{\mu_2 p+\mu_1 p^\prime},~~r_{22}(0)=\frac{\epsilon_2p-\epsilon_1 p^\prime}{\epsilon_2 p+\epsilon_1 p^\prime},\nonumber\\
&&t_{11}(0)=\frac{2\mu_2p}{\mu_2 p+\mu_1 p^\prime},~~
t_{22}(0)=\frac{2\epsilon_2p}{\epsilon_2 p+\epsilon_1 p^\prime},\nonumber\\
&&r_{12}(0)=r_{21}(0)=t_{12}(0)=t_{21}(0)=0,
\end{eqnarray}
where $p$ and $p^\prime$ satisfy (\ref{eq:pqr}).
These conditions can equivalently be obtained from the Fresnel formulas.

The invariant imbedding method can also be used in calculating the wave function $\Psi(z;L)$
inside the inhomogeneous medium.
It turns that the equation satisfied by $\Psi(z;L)$ is very similar to that for $t$ and takes the form
\begin{eqnarray}
&&\frac{\partial}{\partial l}\Psi(z;l)=ip\Psi(z;L){\tilde\mathcal{E}}(l)\nonumber\\&&
~~~~~~~~+\frac{ip}{2}\Psi(z;l)\left\{{\tilde\mathcal{M}}(l)-{\tilde\mathcal{E}}(l)+\frac{q^2}{p^2}\left[{\tilde{\mathcal M}}(l)-{\tilde{\mathcal E}}^{-1}(l)\right]\right\}\left[r(l)+I\right].
\label{eq:fe}
\end{eqnarray}
This equation is integrated from $l=z$ to $l=L$ using the initial condition $\Psi(z;z)=I+r(z)$
to obtain $\Psi(z;L)$.

\section{Conservation of energy}
\label{sec4}

In our formalism, the reflection and transmission coefficients are defined in terms of both the electric and magnetic fields.
For example, in the definition of $r_{21}$, the incident wave is $E_y$ and the reflected wave is
$H_y$. In uniform dielectric media, the magnitudes of the electric and magnetic fields, $E$ and $H$, associated with a plane wave are related to each other by $E=\eta H$, where $\eta$ ($=\sqrt{\mu/\epsilon}$) is the wave impedance.
Therefore it is necessary to
take this factor into account to compare our coefficients with those of other researchers defined using only electric fields.
Specifically, we compare our definitions of $r$ and $t$ with the reflection and transmission coefficients $r_{ss}$, $r_{sp}$, $r_{ps}$,
$r_{pp}$, $t_{ss}$, $t_{sp}$, $t_{ps}$ and $t_{pp}$ given in \cite{4} for isotropic chiral media and obtain the correspondences
\begin{eqnarray}
&&r_{11}=r_{ss},~~r_{12}=\eta_1 r_{ps},~~r_{21}=-\frac{1}{\eta_1} r_{sp},~~r_{22}=-r_{pp},\nonumber\\
&&t_{11}=t_{ss},~~t_{12}=\eta_1 t_{ps},~~t_{21}=\frac{1}{\eta_2} t_{sp},~~t_{22}=\frac{\eta_1}{\eta_2}t_{pp},
\label{eq:lek}
\end{eqnarray}
where $\eta_1$ ($=\sqrt{\mu_1/\epsilon_1}$) and $\eta_2$ ($=\sqrt{\mu_2/\epsilon_2}$) are the wave impedances in the incident and transmitted regions.

In the absence of dissipation, the reflected plus transmitted fluxes of energy must be equal to the incident flux.
In case $p^\prime$ is real, the law of energy conservation is expressed as
\begin{eqnarray}
&&\frac{p}{\mu_1}\left(1-\vert r_{ss}\vert^2-\vert r_{sp}\vert^2\right)=\frac{p^\prime}{\mu_2}\left(\vert t_{ss}\vert^2+\vert t_{sp}\vert^2\right),
\nonumber\\
&&\frac{p}{\mu_1}\left(1-\vert r_{ps}\vert^2-\vert r_{pp}\vert^2\right)=\frac{p^\prime}{\mu_2}\left(\vert t_{ps}\vert^2+\vert t_{pp}\vert^2\right),
\end{eqnarray}
for incident $s$ and $p$ polarizations respectively.
If $p^\prime$ is imaginary, there is only exponentially-damped penetration and no transmitted energy flux into the medium at $z<0$ and the left-hand sides of the above equations vanish. The meaning of the transmission coefficients $t_{ij}$'s changes to that of
the coefficients of exponentially-decaying waves \cite{29}. When $p^\prime$ is real and
there is dissipation, the absorptances $\mathcal A_s$ and $\mathcal A_p$ for
$s$ and $p$ waves, which are the fractions of the incident wave energy absorbed into the medium, can be defined as
\begin{eqnarray}
{\mathcal A_s}&=&1-\vert r_{ss}\vert^2-\vert r_{sp}\vert^2-\frac{\mu_1p^\prime}{\mu_2p}\left(\vert t_{ss}\vert^2+\vert t_{sp}\vert^2\right)
\nonumber\\
&=&1-\vert r_{11}\vert^2-\vert \eta_1 r_{21}\vert^2\nonumber\\&&-\frac{\mu_1{\left(\epsilon_2\mu_2-\epsilon_1\mu_1\sin^2\theta\right)}^{1/2}}{\mu_2\sqrt{\epsilon_1\mu_1}\cos\theta}\left(\vert t_{11}\vert^2+\vert \eta_2 t_{21}\vert^2\right),\nonumber\\
{\mathcal A_p}&=&1-\vert r_{ps}\vert^2-\vert r_{pp}\vert^2-\frac{\mu_1p^\prime}{\mu_2p}\left(\vert t_{ps}\vert^2+\vert t_{pp}\vert^2\right)\nonumber\\
&=&1-\bigg\vert \frac{1}{\eta_1}r_{12}\bigg\vert^2-\vert r_{22}\vert^2\nonumber\\&&-\frac{\mu_1{\left(\epsilon_2\mu_2-\epsilon_1\mu_1\sin^2\theta\right)}^{1/2}}
{\mu_2\sqrt{\epsilon_1\mu_1}\cos\theta}\left(\bigg\vert \frac{1}{\eta_1}t_{12}\bigg\vert^2+\bigg\vert \frac{\eta_2}{\eta_1}t_{22}\bigg\vert^2\right).
\end{eqnarray}
If $p^\prime$ is imaginary, ${\mathcal A_s}$ and ${\mathcal A_p}$ are given by
\begin{eqnarray}
{\mathcal A_s}=1-\vert r_{11}\vert^2-\vert \eta_1 r_{21}\vert^2,~~~
{\mathcal A_p}
=1-\bigg\vert \frac{1}{\eta_1}r_{12}\bigg\vert^2-\vert r_{22}\vert^2.
\end{eqnarray}

For circularly-polarized incident waves, we introduce a new set of the reflection and transmission coefficients $r_{ij}$ and $t_{ij}$, where
$i$ and $j$ are either $+$ or $-$. The symbol $+$ ($-$) represents right (left) circular polarization.
For instance, $r_{-+}$ represents the reflection
coefficient when the incident wave is RCP
and the reflected wave is LCP.
Following \cite{4} and using (\ref{eq:lek}), we express these coefficients as
\begin{eqnarray}
r_{++}&=&\frac{1}{2}\left(r_{11}+r_{22}\right)+\frac{i}{2}\left(\eta_1r_{21}-\frac{1}{\eta_1}r_{12}\right),\nonumber\\
r_{+-}&=&-\frac{1}{2}\left(r_{11}-r_{22}\right)+\frac{i}{2}\left(\eta_1r_{21}+\frac{1}{\eta_1}r_{12}\right),\nonumber\\
r_{-+}&=&-\frac{1}{2}\left(r_{11}-r_{22}\right)-\frac{i}{2}\left(\eta_1r_{21}+\frac{1}{\eta_1}r_{12}\right),\nonumber\\
r_{--}&=&\frac{1}{2}\left(r_{11}+r_{22}\right)-\frac{i}{2}\left(\eta_1r_{21}-\frac{1}{\eta_1}r_{12}\right),\nonumber\\
t_{++}&=&\frac{1}{2}\left(t_{11}+\frac{\eta_2}{\eta_1}t_{22}\right)+\frac{i}{2}\left(\eta_2t_{21}-\frac{1}{\eta_1}t_{12}\right),\nonumber\\
t_{+-}&=&-\frac{1}{2}\left(t_{11}-\frac{\eta_2}{\eta_1}t_{22}\right)+\frac{i}{2}\left(\eta_2t_{21}+\frac{1}{\eta_1}t_{12}\right),\nonumber\\
t_{-+}&=&-\frac{1}{2}\left(t_{11}-\frac{\eta_2}{\eta_1}t_{22}\right)-\frac{i}{2}\left(\eta_2t_{21}+\frac{1}{\eta_1}t_{12}\right),\nonumber\\
t_{--}&=&\frac{1}{2}\left(t_{11}+\frac{\eta_2}{\eta_1}t_{22}\right)-\frac{i}{2}\left(\eta_2t_{21}-\frac{1}{\eta_1}t_{12}\right).
\end{eqnarray}
When $p^\prime$ is real, the absorptances $\mathcal A_+$ and $\mathcal A_-$ for
RCP and LCP waves respectively are defined as
\begin{eqnarray}
{\mathcal A_+}&=&1-\vert r_{++}\vert^2-\vert r_{-+}\vert^2\nonumber\\&&-\frac{\mu_1{\left(\epsilon_2\mu_2-\epsilon_1\mu_1\sin^2\theta\right)}^{1/2}}{\mu_2\sqrt{\epsilon_1\mu_1}\cos\theta}\left(\vert t_{++}\vert^2+\vert t_{-+}\vert^2\right),
\nonumber\\
{\mathcal A_-}&=&1-\vert r_{+-}\vert^2-\vert r_{--}\vert^2\nonumber\\&&-\frac{\mu_1{\left(\epsilon_2\mu_2-\epsilon_1\mu_1\sin^2\theta\right)}^{1/2}}{\mu_2\sqrt{\epsilon_1\mu_1}\cos\theta}\left(\vert t_{+-}\vert^2+\vert t_{--}\vert^2\right).
\end{eqnarray}
Again, if $p^\prime$ is imaginary, the last terms expressed in terms of the transmission coefficients are omitted and ${\mathcal A_+}$ and
${\mathcal A_-}$ are given by
\begin{eqnarray}
{\mathcal A_+}=1-\vert r_{++}\vert^2-\vert r_{-+}\vert^2,~~~
{\mathcal A_-}=1-\vert r_{+-}\vert^2-\vert r_{--}\vert^2.
\end{eqnarray}

\section{Applications}
\label{sec5}

In this section, we apply our method to several examples. The main purpose is not to give a complete analysis of these examples, but to
illustrate the accuracy and usefulness of the method developed in this paper.

\subsection{Uniform chiral slab}

We first consider a uniform slab made of an isotropic chiral medium, where $\epsilon$, $\mu$ and $\gamma$ are constants and $\chi$ is zero.
Plane waves are incident on this slab obliquely from a dielectric with $\epsilon=\epsilon_1$ and $\mu=\mu_1$ and transmitted to another
dielectric with $\epsilon=\epsilon_2$ and $\mu=\mu_2$. The analytical calculation of the matrix reflection and transmission coefficients in this case is complicated but straightforward and Lekner has given a simple prescription of how to do this \cite{4}. We have calculated $r$ and $t$ using both our
method and Lekner's semi-analytical method and confirmed that both methods give identical results.

As an example, we consider the situation where $s$ and $p$ waves of vacuum wavelength $\lambda$ are incident on a uniform chiral slab of thickness $L=5\lambda$ with parameters $\epsilon=5$, $\mu=1$ and $\gamma=0.5$ obliquely from a dielectric with $\epsilon=2$ and $\mu=1$ and transmitted to another dielectric with $\epsilon=3$ and $\mu=1$. In figure 1, we plot the reflectances $\vert r_{11}\vert^2$, $\vert r_{22}\vert^2$ and $\vert r_{12}\vert^2$ ($=\vert r_{21}\vert^2$) obtained using our method and the method of \cite{4} versus incident angle. The agreement is perfect. The transmission coefficients obtained using the two methods also show a perfect agreement.

\begin{figure}
\centering\includegraphics[width=8cm]{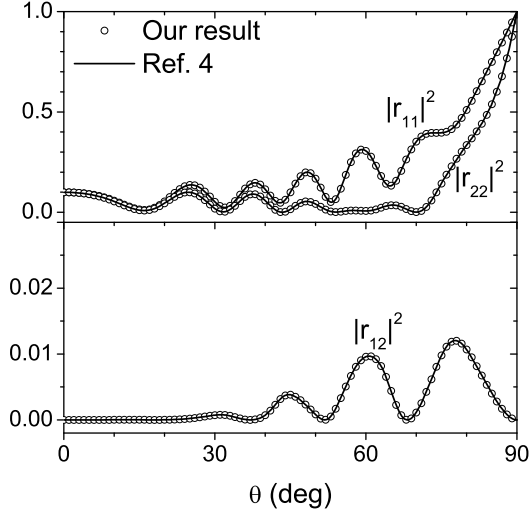}
\caption{Reflectances $\vert r_{11}\vert^2$, $\vert r_{22}\vert^2$ and $\vert r_{12}\vert^2$ ($=\vert r_{21}\vert^2$) obtained using our method and the method of \cite{4} plotted versus incident angle. Plane waves of vacuum wavelength $\lambda$ are incident on a uniform chiral slab of thickness $L=5\lambda$ with parameters $\epsilon=5$, $\mu=1$ and $\gamma=0.5$ obliquely from a dielectric with $\epsilon=2$ and $\mu=1$ and transmitted to another dielectric with $\epsilon=3$ and $\mu=1$.}
\end{figure}

\subsection{Surface waves in Tellegen media}

In this subsection, we illustrate how to use our method to investigate the surface waves excited at the interface between a bi-isotropic medium and
an ordinary dielectric or a metal. We consider a bilayer system consisting of a layer of Tellegen medium with $\epsilon=2.13$, $\mu=1$ and $\chi\ne 0$ and a silver layer with $\epsilon=-16+i$ and $\mu=1$. The bi-isotropic parameter of naturally-occurring materials is usually quite small. During the last decade, however, there has been much progress in fabricating artificial metamaterials showing large values of the magnetoelectric parameter \cite{30,31}. Plane waves of vacuum wavelength $\lambda=622$ nm are incident from a prism with $\epsilon=3.13$ and $\mu=1$ on the Tellegen layer side of the bilayer system and transmitted to a vacuum region. The thicknesses of the Tellegen layer and the metal layer are 320 nm and 150 nm respectively.

In figure 2, we plot the absorptances ${\mathcal A}_s$ and ${\mathcal A}_p$ for $s$ and $p$ waves versus incident angle when $\chi=0.4$ and 0.8 and compare the results with the case where the Tellegen layer is replaced by an ordinary dielectric layer with $\chi=0$. When $\chi$ is zero, there appears a peak at $\theta\approx 62.4^\circ$ only for incident $p$ waves, which corresponds to the ordinary surface plasmon. When $\chi$ is nonzero, both curves show a peak at the same angle, which is $\theta\approx 58.0^\circ$ for $\chi=0.4$ and $\theta\approx 46.6^\circ$ for $\chi=0.8$. That this is due to the excitation of a surface wave can be verified directly by solving the analytical dispersion relation for surface waves between a semi-infinite Tellegen medium with $\epsilon=\epsilon_1$, $\mu=\mu_1$ and $\chi=\chi_1$ and
a semi-infinite dielectric with $\epsilon=\epsilon_2$ and $\mu=\mu_2$, which has the form
\begin{eqnarray}
&&\left(\epsilon_2\kappa_1+\epsilon_1\kappa_2\right)\left(\mu_2\kappa_1+\mu_1\kappa_2\right)-\kappa_2^2\chi_1^2=0,\nonumber\\
&&\kappa_1=\sqrt{\beta^2-(\epsilon_1\mu_1-\chi_1^2)},~~\kappa_2=\sqrt{\beta^2-\epsilon_2\mu_2},
\label{eq:sw}
\end{eqnarray}
where $\beta$ is the magnitude of the wave vector component parallel to the interface normalized by the vacuum wave number.
By numerically solving (\ref{eq:sw}) after substituting $\epsilon_1=2.13$, $\mu_1=1$, $\chi_1=0.4$ (0.8), $\epsilon_2=-16+i$, $\mu_2=1$ and $\beta=\sqrt{3.13}\sin\theta$ into it, we find that surface waves are excited at $\theta\approx 58.0^\circ$ ($46.6^\circ$), which agrees with our result shown in figure 2 precisely. We notice that in Tellegen media, surface waves are excited for both $s$ and $p$ waves. The invariant imbedding method can be applied equally easily to the investigation of surface waves in more complicated situations, in which $\epsilon$, $\mu$, $\chi$ and $\gamma$ vary arbitrarily along one spatial direction \cite{32}.

\begin{figure}
\centering\includegraphics[width=8cm]{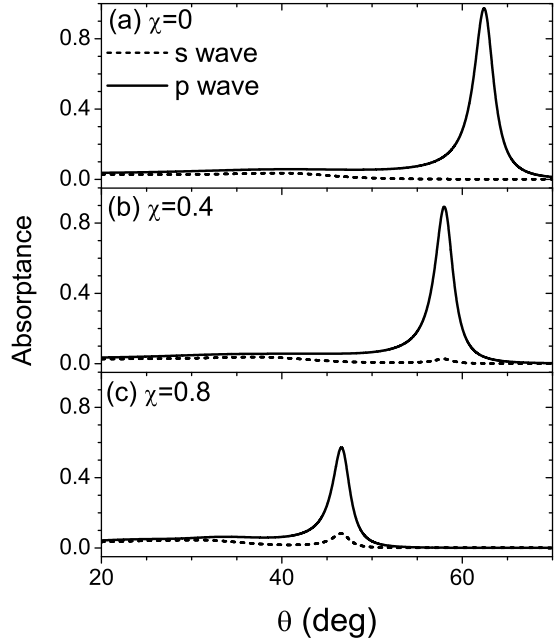}
\caption{Absorptances ${\mathcal A}_s$ and ${\mathcal A}_p$ for $s$ and $p$ waves incident on a bilayer system consisting of a layer of Tellegen medium with $\epsilon=2.13$, $\mu=1$ and $\chi=0$, 0.4, 0.8 and a silver layer with $\epsilon=-16+i$ and $\mu=1$ plotted versus incident angle. Plane waves of vacuum wavelength $\lambda=622$ nm are incident from a prism with $\epsilon=3.13$ and $\mu=1$ on the Tellegen layer side of the bilayer system and transmitted to a vacuum region. The thicknesses of the Tellegen layer and the silver layer are 320 nm and 150 nm respectively.}
\end{figure}

In figure 3, we plot the incident angle at which the surface wave is excited, $\theta_s$, for the same configuration of prism/Tellegen layer/metal layer considered in figure 2 as a function of the Tellegen parameter $\chi$ of the Tellegen layer. The result obtained using (32) is compared with that obtained from the invariant imbedding method and the agreement is seen to be excellent. We observe that $\theta_s$ decreases to smaller angles as $\chi$ increases.

\begin{figure}
\centering\includegraphics[width=8cm]{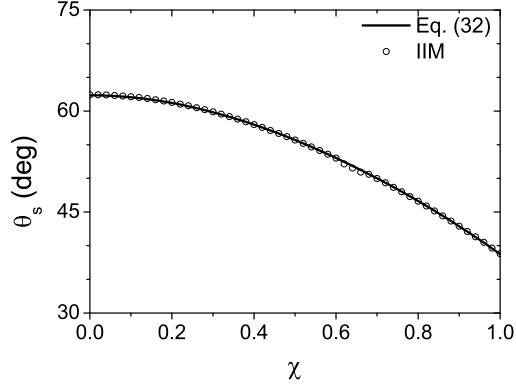}
\caption{Incident angle at which the surface wave is excited, $\theta_s$, for the configuration of prism/Tellegen layer/metal layer considered in figure 2 plotted versus the Tellegen parameter $\chi$ of the Tellegen layer. The result obtained using the analytical dispersion relation (32) is compared with that obtained from the invariant imbedding method.}
\end{figure}

\subsection{Mode conversion in inhomogeneous Tellegen media}

As a third example, we consider the mode conversion in inhomogeneous Tellegen media. Mode conversion is a phenomenon where transverse electromagnetic waves are resonantly converted to longitudinal plasma oscillations \cite{21,22,33,34,35}. In ordinary unmagnetized plasmas, this occurs at the regions where the local refractive index vanishes. In Tellegen media, the effective refractive index squared is given by $\epsilon\mu-\chi^2$. If there exists a region
where the real part of this quantity vanishes inside an inhomogeneous medium, the mode conversion and the associated resonant absorption occur for both incident $s$ and $p$ waves. The absorptance, which can equivalently be called the mode conversion coefficient, is finite even in the presence of an infinitesimally small amount of damping at the resonance region,
which signifies that the absorption is not due to any dissipative damping process but due to mode conversion.

In figure 4, we plot the absorptances ${\mathcal A}_s$ and ${\mathcal A}_p$ for $s$ and $p$ waves incident on a nonuniform Tellegen layer of thickness $L$ versus incident angle. The dielectric permittivity of the Tellegen medium is given by $\epsilon=\epsilon_R+i\epsilon_I$, where $\epsilon_R=z/L+0.2$ ($0\le z\le L$) and  $\epsilon_I=10^{-8}$. The other parameters used are $\mu=1$, $\chi=0$, 0.5, 0.7, 1.1 and $k_0L=10\pi$. When $\chi$ is equal to 0.5 and 0.7, the real part of the effective refractive index vanishes at $z=0.05L$ and $z=0.29L$ respectively. We find that substantial mode conversion occurs for both polarizations when $\theta<30^\circ$, though it is stronger for $p$ waves than for $s$ waves.
On the other hand, when $\chi$ is equal to 0 and 1.1, the real part of the effective refractive index never vanishes in the region $0\le z\le L$ 
and no mode conversion occurs. 
Again, our method can be easily adapted to the more general cases where $\epsilon$, $\mu$, $\chi$ and $\gamma$ vary arbitrarily along one spatial direction.

\begin{figure}
\centering\includegraphics[width=8cm]{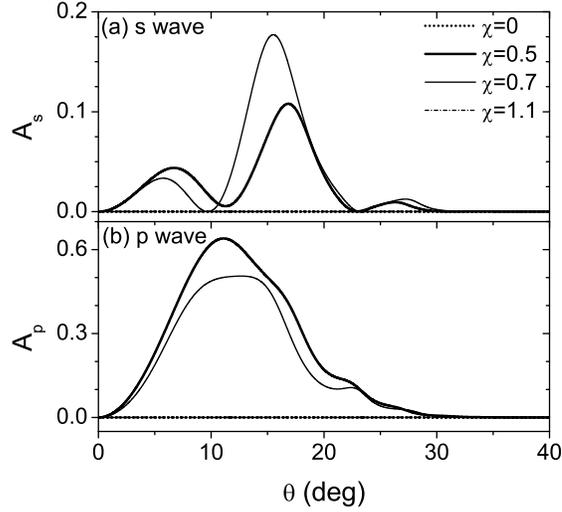}
\caption{Absorptances or, equivalently, mode conversion coefficients ${\mathcal A}_s$ and ${\mathcal A}_p$ for $s$ and $p$ waves incident on a nonuniform Tellegen layer of thickness $L$ plotted versus incident angle, when $\epsilon_R=z/L+0.2$ ($0\le z\le L$), $\epsilon_I=10^{-8}$, $\mu=1$, $\chi=0$, 0.5, 0.7, 1.1 and $k_0L=10\pi$.}
\end{figure}

\section{Conclusion}
\label{sec6}

In this paper, we have studied theoretically the wave propagation in general bi-isotropic media, which include isotropic chiral media and Tellegen media as special cases. We have developed a new generalized version of the invariant imbedding method, which can be used to calculate wave propagation characteristics of coupled waves in arbitrarily-inhomogeneous stratified bi-isotropic media accurately and efficiently. The invariant imbedding equations we have derived are novel and versatile and can be applied to a wide variety of problems involving bi-isotropic media. We have applied our method to several examples, including the surface wave excitation at the interface between a Tellegen medium and a metal and the mode conversion of transverse electromagnetic waves into longitudinal plasma oscillations occurring in inhomogeneous Tellegen media. From these examples, we have verified the validity of our method and demonstrated its usefulness and efficiency. We expect our method to be quite useful in exploring and analyzing various kinds of new physical phenomena involving chiral media, Tellegen media and general bi-isotropic media, since it is easy to use, efficient and accurate. Because topological insulators can be considered to be a kind of Tellegen media, it will be also useful in studying those interesting materials.

\ack
This work has been supported by the National Research Foundation of Korea Grant (NRF-2015R1A2A2A01003494) funded by the Korean Government.

\appendix
\section{Invariant imbedding method}
\label{sec-imbed}

The invariant imbedding equations in very general cases have been derived a long time ago by Golberg \cite{27}.
Here we follow the presentation given in \cite{28} closely.
We consider a boundary value problem of an arbitrary number of coupled first-order ordinary differential equations defined by
\begin{eqnarray}
&&\frac{d}{dz} {\bf u}(z)={\bf F}(z,{\bf u}(z)),~~ z \in [0,L],\label{eq:uu} \\
&&g{\bf u}(0)+h{\bf u}(L)={\bf v},\label{eq:bc}
\end{eqnarray}
where $\bf u$, $\bf F$ and $\bf v$ are $N$-component vectors and $g$ and $h$ are constant $N\times N$ matrices.
We consider the function $\bf u$ as being parametrically dependent on $L$ and $\bf v$:
\begin{eqnarray}
{\bf u}(z)={\bf u}(z;L,{\bf v})
\end{eqnarray}
and introduce
\begin{eqnarray}
{\bf R}(L,{\bf v})={\bf u}(L;L,{\bf v}),~~~ {\bf S}(L,{\bf v})={\bf u}(0;L,{\bf v}).
\end{eqnarray}
Next we take derivatives of (\ref{eq:uu}) with respect to $L$ and $\bf v$ respectively and obtain
\begin{eqnarray}
\frac{d}{dz}\frac{\partial u_i(z;L,{\bf v})}{\partial L}&=&\frac{\partial F_i(z,{\bf u})}{\partial u_j}\frac{\partial u_j(z;L,{\bf v})}{\partial L},  \nonumber\\
\frac{d}{dz}\frac{\partial u_i(z;L,{\bf v})}{\partial v_k}&=&\frac{\partial F_i(z,{\bf u})}{\partial u_j}\frac{\partial u_j(z;L,{\bf v})}{\partial v_k},
\end{eqnarray}
where the summation over repeated indices is implied.
These equations are identical in form and their solutions are related by the linear expression
\begin{eqnarray}
\frac{\partial u_i(z;L,{\bf v})}{\partial L}=\lambda_k(L,{\bf v}) \frac{\partial u_i(z;L,{\bf v})}{\partial v_k},
\label{eq:lam}
\end{eqnarray}
where the vector $\mathbf \lambda$ needs to be specified.
From (\ref{eq:lam}), we obtain
\begin{eqnarray}
&&g\frac{\partial {\bf u}(0;L,{\bf v})}{\partial L}+h\frac{\partial {\bf u}(z;L,{\bf v})}{\partial L}\bigg\vert_{z=L}\nonumber \\
&&~~~=\lambda_k(L,{\bf v})\frac{\partial}{\partial v_k}\left[g{\bf u}(0;L,{\bf v})+h{\bf u}(L;L,{\bf v})\right] \nonumber \\
&&~~~=\lambda_k(L,{\bf v})\frac{\partial {\bf v}}{\partial v_k}={\mathbf{\lambda}}(L,{\bf v}).
\end{eqnarray}
Since
\begin{eqnarray}
&&\frac{\partial {\bf u}(z;L,{\bf v})}{\partial L}\bigg\vert_{z=L}=\frac{\partial {\bf u}(L;L,{\bf v})}{\partial L}-\frac{\partial {\bf u}(z;L,{\bf v})}{\partial z}\bigg\vert_{z=L} \nonumber \\
&&~~~=\frac{\partial {\bf R}(L,{\bf v})}{\partial L}-{\bf F}(L,{\bf R}(L,{\bf v})),
\end{eqnarray}
we obtain
\begin{eqnarray}
{{\mathbf \lambda}(L,{\bf v})}=-h{\bf F}(L,{\bf R}(L,{\bf v}))
\end{eqnarray}
using (\ref{eq:bc}).
This gives us the desired invariant imbedding equation for the fields in inhomogeneous media:
\begin{eqnarray}
\frac{\partial u_i(z;l,{\bf v})}{\partial l}=-h_{kj}F_j(l,{\bf R}(l,{\bf v}))\frac{\partial u_i(z;l,{\bf v})}{\partial v_k},
\label{eq:iif}
\end{eqnarray}
which is integrated from $l=z$ to $l=L$
using the initial condition
\begin{equation}
{\bf u}(z;z,{\bf v})={\bf R}(z,{\bf v}).
\end{equation}

In order to solve (\ref{eq:iif}), we need to calculate the function ${\bf R}(l,{\bf v})$ for $0\le l \le L$ in advance.
From the identity
\begin{eqnarray}
\frac{\partial {\bf R}(L,{\bf v})}{\partial L}
=\frac{\partial {\bf u}(z;L,{\bf v})}{\partial z}\bigg\vert_{z=L}+\frac{\partial {\bf u}(z;L,{\bf v})}{\partial L}\bigg\vert_{z=L},
\end{eqnarray}
we get the invariant imbedding equation for $\bf R$:
\begin{eqnarray}
\frac{\partial {\bf R}(l,{\bf v})}{\partial l}
&=&-h_{kj}F_j(l,{\bf R}(l,{\bf v}))\frac{\partial {\bf R}(l,{\bf v})}{\partial v_k}\nonumber\\
&&+{\bf F}(l,{\bf R}(l,{\bf v})),
\label{eq:iir}
\end{eqnarray}
which is integrated from $l=0$ to $l=L$
using the initial condition
\begin{equation}
{\bf R}(0,{\bf v})=(g+h)^{-1}{\bf v}.
\end{equation}
In a similar manner, we obtain the invariant imbedding equation for $\bf S$:
\begin{eqnarray}
\frac{\partial {\bf S}(l,{\bf v})}{\partial l}=-h_{kj}F_j(l,{\bf R}(l,{\bf v}))\frac{\partial {\bf S}(l,{\bf v})}{\partial v_k}
\label{eq:iis}
\end{eqnarray}
with the initial condition
\begin{equation}
{\bf S}(0,{\bf v})=(g+h)^{-1}{\bf v}.
\end{equation}
This equation can be solved together with (\ref{eq:iir}).

Next we specialize to the case where the function ${\bf F}(z,{\bf u})$ is linear in $\bf u$ such that
\begin{eqnarray}
F_i(z,{\bf u}(z))=A_{ij}(z)u_j(z).
\end{eqnarray}
Then the function ${\bf u}$ is linear in $\bf v$:
\begin{eqnarray}
{\bf u}(z;l,{\bf v})={\mathcal U}(z;l){\bf v},
\end{eqnarray}
where $\mathcal U$ is an $N\times N$ matrix.
Substituting this into (\ref{eq:iif}), we obtain a differential equation for the matrix function $\mathcal U$:
\begin{eqnarray}
\frac{\partial}{\partial l}{\mathcal U}(z;l)=-{\mathcal U}(z;l)hA(l){\mathcal R}(l)
\end{eqnarray}
supplemented by the initial condition
\begin{eqnarray}
{\mathcal U}(z;z)={\mathcal R}(z),
\end{eqnarray}
where we have used the definition ${\bf R}(l,{\bf v})={\mathcal R}(l){\bf v}$.
The invariant imbedding equations for the matrices ${\mathcal R}$ and $\mathcal S$, which is defined by ${\bf S}(l,{\bf v})={\mathcal S}(l){\bf v}$, are similarly obtained:
\begin{eqnarray}
\frac{d}{dl}{\mathcal R}(l)&=&A(l){\mathcal R}(l)-{\mathcal R}(l)h A(l){\mathcal R}(l),\nonumber\\
\frac{d}{dl}{\mathcal S}(l)&=&-{\mathcal S}(l)h A(l){\mathcal R}(l),
\label{eq:imbed1}
\end{eqnarray}
together with the initial conditions
\begin{eqnarray}
{\mathcal R}(0)={\mathcal S}(0)=(g+h)^{-1}.
\label{eq:ic}
\end{eqnarray}


\section*{References}
\begin{thebibliography}{10}

\bibitem{1} Lindell I V, Sihvola A H, Tretyakov S A and Viitanen A J 1994 {\it Electromagnetic Waves
in Chiral and Bi-Isotropic Media} (Norwood, MA: Artech House)

\bibitem{2} Fiebig M 2005 Revival of the magnetoelectric effect {\it J. Phys. D} {\bf 38} R123

\bibitem{3} Pyatakov A P and Zvezdin A K 2012 Magnetoelectric and multiferroic media {\it Phys. Usp.} {\bf 55} 557

\bibitem{4} Lekner J 1996 Optical properties of isotropic chiral media {\it Pure Appl. Opt.}
{\bf 5} 417

\bibitem{5} Obukhov Y N and Hehl F W 2009 On the boundary-value problems and the validity of the Post constraint in modern electromagnetism {\it Optik} {\bf 120} 418

\bibitem{6} Kamenetskii E O, Sigalov M and Shavit R 2009 Tellegen particles and magnetoelectric metamaterials {\it J. Appl. Phys.} {\bf 105} 013537
\bibitem{7} Prud\^encio F R, Matos S A and Paiva C R 2014 A geometrical approach to duality transformations for Tellegen media {\it IEEE Trans. Microw. Theory Tech.} {\bf 62} 1417

\bibitem{8} Wang B, Zhou J, Koschny T,
Kafesaki M and Soukoulis C M 2009 Chiral metamaterials: simulations and
experiments {\it J. Opt. A: Pure Appl. Opt.} {\bf 11} 114003

\bibitem{9} Plum E, Zhou J, Dong J, Fedotov V A, Koschny T, Soukoulis C M and Zheludev N I 2009
Metamaterial with negative index due to chirality {\it Phys. Rev. B} {\bf 79} 035407

\bibitem{10} Zhang S, Park Y, Li J, Lu X, Zhang W and Zhang X 2009 Negative refractive index in chiral metamaterials {\it Phys. Rev. Lett.} {\bf 102} 023901

\bibitem{11} Li Z, Mutlu M and Ozbay E 2013 Chiral metamaterials: from optical
activity and negative refractive index to
asymmetric transmission {\it J. Opt.} {\bf 15} 023001

\bibitem{12} Kim K, Yoo H and Lim H 2006 Exact analytical expressions for the dispersion relation
of one-dimensional chiral photonic crystals {\it Waves Random Complex
Media} {\bf 16} 75

\bibitem{13} Lee K J, Wu J W and Kim K 2014 Defect modes in a one-dimensional photonic crystal with a chiral defect layer {\it Opt. Mater. Express} {\bf 4} 2542

\bibitem{14} Hasan M Z and Kane C L 2010 Colloquium: topological insulators {\it Rev. Mod. Phys.} {\bf 82} 3045

\bibitem{15} Qi X and Zhang S 2011 Topological insulators and superconductors {\it Rev. Mod. Phys.} {\bf 83} 1057

\bibitem{16} Tse W and MacDonald A H 2010 Giant magneto-optical Kerr effect and universal Faraday effect in thin-film topological insulators {\it Phys. Rev. Lett.} {\bf 105} 057401

\bibitem{17} Karch A 2011 Surface plasmons and topological insulators {\it Phys. Rev. B} {\bf 83} 245432

\bibitem{18} Li L L and Xu W 2014 Surface plasmon polaritons in a topological insulator embedded in an optical cavity {\it Appl. Phys. Lett.} {\bf 104} 111603

\bibitem{19}
Klyatskin V I 1994 The imbedding method in statistical
boundary-value wave problems {\it Prog. Opt.} {\bf 33} 1

\bibitem{20} Kim K, Lim H and Lee D 2001 Invariant imbedding
equations for electromagnetic waves in stratified magnetic media:
Applications to one-dimensional photonic crystals {\it J. Korean Phys.
Soc.} {\bf 39} L956

\bibitem{21}
Kim K and Lee D 2005 Invariant imbedding theory of mode
conversion in inhomogeneous plasmas. I. Exact calculation of the
mode conversion coefficient in cold, unmagnetized plasmas {\it Phys.
Plasmas} {\bf 12} 062101

\bibitem{22} Kim K and Lee D 2006 Invariant imbedding theory of mode conversion in inhomogeneous
plasmas. II. Mode conversion in cold, magnetized plasmas with
perpendicular inhomogeneity {\it Phys. Plasmas} {\bf 13} 042103

\bibitem{23}
Kim K, Phung D K, Rotermund F and Lim H 2008 Propagation of
electromagnetic waves in stratified media with nonlinearity in both
dielectric and magnetic responses {\it Opt. Express} {\bf 16}
1150

\bibitem{24} Kim K, Lee D and Lim H 2005 Theory of the propagation of coupled waves in
arbitrarily inhomogeneous stratified media {\it EPL} {\bf
69} 207

\bibitem{25} Serdyukov A, Semchenko I, Tretyakov S and Sihvola A 2001 {\it Electromagnetics of Bi-Anisotropic Materials} (Amsterdam: Gordon and Breach) pp~22--27

\bibitem{26} Landau L D, Lifshitz E M and Pitaevskii L P 1984 {\it Electrodyanmics of Continuous Media} 2nd edn (Oxford: Butterworth-Heinemann) p~358

\bibitem{27} Golberg M A 1971 A generalized invariant imbedding equation {\it J. Math. Anal. Appl.} {\bf 33} 518

\bibitem{28} Klyatskin V I 2005 {\it Stochastic Equations through the Eye of the Physicist} (Amsterdam: Elsevier) pp~440--442 

\bibitem{29} Lekner J 2016 {\it Theory of Reflection} 2nd edn (Berlin: Springer) p~44

\bibitem{30} Tretyakov S A, Maslovski S I, Nefedov I S, Viitanen A J, Belov P A and Sanmartin A 2003 Artificial tellegen particle {\it Electromagnetics} {\bf 23} 665

\bibitem{31} Zhang S, Park Y S, Li J, Lu X, Zhang W and Zhang X 2009 Negative refractive index in chiral metamaterials {\it Phys. Rev. Lett.} {\bf 102} 023901

\bibitem{32} Kim K 2008 Excitation of s-polarized surface electromagnetic waves in inhomogeneous dielectric media {\it Opt. Express} {\bf 16} 13354

\bibitem{33} Swanson D G 1998
{\it Theory of Mode Conversion and Tunneling in Inhomogeneous
Plasmas} (New York: Wiley)

\bibitem{34} Kim K, Lee D and Lim H 2008 Resonant absorption and mode conversion in a transition layer between
positive-index and negative-index media {\it Opt. Express} {\bf 16} 18505

\bibitem{35} Kim S and Kim K 2016 Resonant absorption and amplification
of circularly-polarized waves in
inhomogeneous chiral media {\it Opt. Express} {\bf 24} 1794

\end{thebibliography}
\end{document}